\documentclass[11pt]{article}
\usepackage[margin=1in]{geometry}
\usepackage{hyperref}
\usepackage{xcolor}
\usepackage{amsmath}
\usepackage{enumerate}
\usepackage{graphicx}
\usepackage[backend=bibtex,style=nature]{biblatex}

\bibliography{references}

\title{\Large \textbf{Propensity Score Weighting to Ensure Balance in Key Subgroups or Strata: A Practical Guide}\\ }
\author{Emma K. Mackay$^{1,2}$\thanks{Corresponding author; e-mail: \href{mailto:emma.mackay@unityhealth.to}{\textcolor{blue}{emmak.mackay@utoronto.ca}}}, Amol A. Verma$^{2,3,4}$, Fahad Razak$^{2,3,4}$, Surain B. Roberts$^{2,3}$\\ \; \\ \footnotesize $^1$IQVIA, Toronto, Canada\vspace*{0.02in}\\ \footnotesize $^2$Institute of Health Policy, Management and Evaluation, University of Toronto, Toronto, Canada\vspace*{0.02in}\\ \footnotesize $^3$Li Ka Shing Knowledge Institute, St. Michael's Hospital, Unity Health Toronto, Toronto, Canada\vspace*{0.02in}\\ \footnotesize $^4$Department of Medicine, University of Toronto, Toronto, Canada}
\date{\; \\ \; \\ \today}

\begin{document}
\maketitle
\begin{abstract}
Propensity score weighting approaches have been widely implemented in clinical research to estimate the effects of a treatment or exposure while mitigating the risk of confounding in the absence of random assignment. Propensity score weighting is used to create a weighted pseudo-population of patients in which observed baseline characteristics (i.e. potential confounders) are balanced between exposure groups. If all measured and unmeasured confounders are balanced between exposure groups in the pseudo-population, any differences in outcomes by exposure group can be attributed to the treatment received. Assuming adequate balance is achieved, we can compute estimates of the effect of exposure by comparing outcomes between exposure groups in the weighted pseudo-population. 

In practice, when working with large electronic health records (EHR) or administrative datasets to evaluate health quality outcomes at the institutional level, or evaluate supportive care interventions for a wide range of hospitalized patients, it may be advisable to stratify the propensity score weighting approach by indication, reason for admission, or other clinical risk factors due to the potential for substantial heterogeneity across subgroups of patients with complex care needs. 

A stratified approach may be appropriate if (i) prognosis differs substantially between patient subgroups such that achieving balance in the composition of these strata between exposure/treatment groups should be prioritized, (ii) likelihood of exposure differs substantially across clinical subgroups, or (iii) the covariate-exposure associations are expected to differ substantially between subgroups (i.e. there are covariate-subgroup interactions in the exposure/treatment propensity model). For example, we may want to evaluate the impact of prophylactic anticoagulant use for venous thromboembolism prevention in elderly patients admitted to hospital for a wide array of conditions.

The purpose of this article is to outline an approach to implementing propensity score weighting with stratification by clinical groups. We also provide guidance on best practices with particular focus on EHR and administrative medical data, and population health settings.
\end{abstract}
\vspace*{0.2in}
\textbf{Keywords:} Propensity scores, propensity score weighting, inverse probability of treatment weighting, causal inference, observational data, stratification
\newpage

\section{Background}

There is substantial interest from researchers, regulators, and health technology assessment bodies in the use of `real-world' or `observational' data collected from electronic health records, insurance claims, or administrative data sources to infer the causal effects of different treatments or `exposures'\cite{hennessy2025,patel2021,hernan2016,hernan2022}. However, in the absence of randomization to the exposure, effect estimates may be biased due to selection into treatment--with differences in baseline prognosis between exposure groups incorrectly attributed the effect of exposure.

To mitigate the risk of confounding in causal inference using observational data, a number of methods can be used, including regression adjustment or `outcome regression' methods such as standardization or parametric g-computation\cite{hernan2020}, matching methods\cite{ho2007}, and weighting methods based on either propensity scores\cite{rosenbaum1983,austin2011,williamson2012} or covariate balancing\cite{zubizarreta2015}. Outcome regression approaches attempt to account for differences in patient characteristics between exposure groups using a parametric modelling approach. Matching and weighting approaches both attempt to construct a (matched or weighted) pseudo-population in which patient characteristics are balanced between exposure groups. The treatment effect estimate under weighting and matching is then computed by comparing outcomes between exposure groups in this balanced pseudo-population. Doubly-robust methods which incorporate both outcome regression and exposure/propensity score modelling and are robust to misspecification of either the outcome model or exposure/propensity score model (but not both) can also be used\cite{funk2011,schuler2017}.  

When medical researchers apply propensity score weighting or matching methods in practice, they often aim to construct a single pseudo-population in which patient baseline characteristics are balanced between the exposed and unexposed groups. However, in many settings the population of interest may consist of extremely heterogeneous subpopulations in terms of both prognosis and propensity to be exposed/treated. In these settings we may want to ensure that we prioritize achieving balance in terms of the subgroup (or 'stratum') composition between exposure groups and to use a propensity score modelling approach that allows for the process of selection into exposure/treatment to differ between strata. 

For example, suppose we would like to evaluate the impact of sedative-hypnotic use as a sleep aid on the risk of falls among elderly hospital inpatients. Patients admitted for diabetes complications are likely to differ from those admitted for fractures in terms of (a) their baseline risk for an in-hospital fall, (b) their likelihood of being given a sedative-hypnotic, and (c) how other factors such as age are likely to influence the decision to prescribe sedative-hypnotics (covariate-subgroup interactions in the exposure model) and their risk of a fall (covariate-subgroup interactions in the outcome model). In this case, rather than fitting a highly saturated propensity score model, we may want to consider fitting separate propensity score models by subgroup and evaluating post-weighting balance within each subgroup. To implement such a stratified approach we would also want to combine together each of these stratum-specific subpopulations into an overall population in which the stratum shares are balanced between exposure groups. Additionally, we may also want to ensure that the stratum shares in our weighted pseudo-population are the same as in the underlying data, or in a target population in which we want to infer treatment/exposure effects. This may be relevant if stratum is an effect modifier--that is, if the effect size differs by stratum.

Just as we might want to stratify either our randomization procedure or analysis by specific clinical factors in a randomized controlled trial (RCT), we may also want to implement a similar stratification procedure when implementing propensity score weighting or matching methods. Stratified randomization in RCTs is primarily used to ensure balance in key prognostic factors when planned sample sizes are modest\cite{cook2007ch1,cook2007ch5}. We may, for example, want to stratify by cancer stage in oncology studies, surgery type in studies of post-surgery pain management, wound care, or prophylactic antibiotic use, or reason for admission in studies assessing institutional practices or supportive care interventions for health quality assessment, among other applications.

We outline a procedure for conducting propensity score weighting that allows for stratification by clinical subgroups. Note that this stratified propensity score weighting procedure should not be confused with the method of stratifying on the propensity score or `propensity score stratification' (see, for example, Austin\cite{austin2011} and Williamson et al.\cite{williamson2012}). Our approach is to estimate the propensity score model and corresponding propensity score weights separately in each stratum. We then apply an appropriate scaling factor to the weights in each stratum before pooling the stratum-specific pseudo-populations into a single overall pseudo-population with the same proportion of patients allocated to each stratum as was the case in the underlying unweighted population. We can then estimate a marginal treatment effect for the overall mix of strata by comparing outcomes in the combined weighted pseudo-population.

\section{Methods}
\subsection{Potential Outcomes Framework and Propensity Score Weighting Overview}

We begin with a brief overview of the potential outcomes framework and propensity score weighting. For a more in-depth treatment we refer the reader to Williamson et al.\cite{williamson2012}, Austin\cite{austin2011} and Hern\'an and Robins\cite{hernan2020}. Under the potential outcomes framework, we can characterize the outcome of a patient, $i\in\{1,...,n\}$, under the scenario where they are unexposed ($Y_{0i}$) or exposed ($Y_{1i}$). While we might want to compute the treatment effect for this individual patient, $Y_{1i}-Y_{0i}$, we only ever observe one of the two potential outcomes for patient $i$ whereas the other is purely a hypothetical counter-factual. We can characterize the patient's observed outcome, $Y_i$, as $Y_i = Z_i Y_{1i} + (1-Z_i)Y_{0i}$ where $Z_i \in \{0,1\}$ is an indicator for whether or not the patient was exposed.

We can arrive at valid estimates of the average treatment effect (ATE) from observational data if a few conditions are met. The first one is the stable unit-treatment value assumption (SUTVA): that a patient's outcomes aren't impacted by the other patients' treatment assignments or outcomes--in other words, there must be no interference between patients. The second is the ignorability assumption: that the potential outcomes are conditionally independent of the treatment assignment, or, more formally, that $\{Y_0,Y_1\}\perp Z|X$. Dropping the $i$ subscripts for simplicity, we can express the ATE under these two assumptions as
\begin{align*}
\text{ATE}&=E[Y_{1}-Y_{0}]\\ &= E[Y_1] - E[Y_0]\\ &= E[Y_1|Z=1,X] - E[Y_0|Z=0,X]\;\;\;\;\;\;\;\;\;\;\;\;\;\text{(by ignorability)}\\ &=E[Y|Z=1,X] - E[Y|Z=0,X]
\end{align*}

\noindent This means that if we can appropriately adjust for $X$ (the set of confounders), we can arrive at valid causal effect estimates. However, this requires us to (i) identify and measure all confounders, and (ii) correctly specify any outcome regression or propensity score models we use to adjust for these confounders. It is important to note that we cannot confirm whether or not these conditions hold in practice--even approximately. In contrast, for RCTs the ignorability assumption will trivially hold due to randomization.

Propensity scores are the probability of assignment to a treatment or exposure conditional on observed covariates\cite{rosenbaum1983}:

\begin{align*}
e(\mathbf{x}_i) = \text{Pr}(Z_i = 1|X_i = \mathbf{x}_i)
\end{align*}
where $Z_i$ is the binary indicator for whether the patient was exposed, and $\mathbf{x}_i$ is a vector of realized covariates for patient $i$. In practice, we often use a logistic regression model to estimate these propensity scores when conditioning on both categorical and continuous covariates:
\begin{align*}
Z_i &\sim \text{Bernoulli}(p_i)\\
\text{logit}(p_i)&=\mathbf{x}_i'\mathbf{\beta}\;\;\;\;,
\end{align*}
where our estimated propensity score is $\hat{e}(\mathbf{x}_i)=\text{logit}^{-1}(\mathbf{x}_i'\mathbf{\hat{\beta}})$.

We need one final assumption to hold to be able to apply propensity score weighting: positivity. The positivity assumption requires that there is non-zero probability of both exposure and non-exposure given each patient $i$'s covariates, $\mathbf{x}_i$. In other words, there must be overlap in the characteristics of patients who are exposed and unexposed. 

We can use propensity score weighting to construct a weighted pseudo-population in which patient baseline characteristics, $\mathbf{X}$, are balanced between exposure groups. We use the following formula to compute our ATE weights for each patient $i$:
\begin{align*}
w_{i}&=\frac{Z_i}{\hat{e}(\mathbf{x}_i)} + \frac{1 - Z_i}{1 - \hat{e}(\mathbf{x}_i)}\;\;\;\;.
\end{align*}

We can then compute our ATE estimate, $\hat{ATE}$, as the difference in outcomes by exposure group in the weighted pseudo-population:

\begin{align*}
\hat{ATE}=\frac{\sum_{i=1}^n Z_i w_i y_i}{\sum_{i=1}^n Z_i w_i} - \frac{\sum_{i=1}^n (1 - Z_i) w_i y_i}{\sum_{i=1}^n (1 - Z_i) w_i}\;\;\;\;,
\end{align*}

\noindent where $y_i$ is the outcome for patient $i$. Or, more generally, we can fit a weighted outcome regression model which regresses our outcomes $\mathbf{y}$ on the exposure indicator $\mathbf{Z}$ in the ATE-weighted pseudo-population. 

We focused above on an ATE estimand, however there are other estimands, such as the average treatment effect on the treated (ATT), that we might also be interested in. The choice of estimand is important due to the potential for effect modification--where the relative benefit of treatment varies with patient characteristics. For example, suppose we are interested in estimating the effect of a new cancer drug versus the current standard of care (SoC) in a population of stage III and stage IV patients. If the relative benefit of the new drug versus SoC differs by stage then the ATE and ATT can differ if stage IV patients are more or less likely to be given the new drug than stage III patients. Suppose further that the drug is only being considered for use for stage III cancers. In this case we would want our estimand to reflect the benefit to stage III patients so it may be the case that neither the ATE nor the ATT would be appropriate. The reader is directed to Austin\cite{austin2011} and Williamson et al.\cite{williamson2012} for more details on other estimands for propensity score methods, Phillippo et al.\cite{phillippo2025} for more discussion on effect modification, and to Remiro-Az\'ocar et al.\cite{remiroazocar2025} and the ICH E9(R1) guidelines\cite{ichE9R1} for a more detailed discussion of estimands.

For the remainder of this paper we will focus our attention on the application to stratified ATE weighting, however, the formulas we outline for the weights can be modified to target estimands other than the ATE. In the next section (Section 2.2) we outline a procedure for stratifying the propensity score weighting by clinical subgroups. We then follow this with a practical demonstration of the approach in Section 3.

\subsection{A Propensity Score Weighting Approach with Stratification by Key Clinical Subgroups}

When applying propensity score weighting it may sometimes be desirable to implement the procedure using a stratified approach across key clinical subgroups rather than including subgroup indicator variables directly in the propensity score model. Several undesirable issues can arise when adjusting for imbalances in key subgroups via the propensity score model rather than stratifying:
\begin{enumerate}[(i)]
\item the weighted pseudo-population will not necessarily yield exact balance in the proportions of patients in each clinically important subgroup due to the need to estimate the propensity scores using a parametric model (e.g. using logistic regression), 
\item Balance may be poor for some prognostic factors within strata even if these variables are balanced in the overall pseudo-population if our propensity score model does not adequately capture stratum-covariate interactions in our propensity score model, and
\item the proportion of weighted patients in key clinical subgroups may differ from the proportions in the underlying unweighted data.
\end{enumerate}

Issues (i) and (ii) are of particular concern if prognosis is expected to differ substantially across subgroups and can threaten the internal validity of effect estimates. Issue (iii) is of concern if the effect of exposure/treatment is expected to differ substantially across subgroups--i.e. if subgroup is an effect modifier. In this case our estimates may not generalize well to the overall patient population--our focus under an ATE estimand--and can threaten the external validity of effect estimates. Lastly, issue (ii) can lead to residual confounding that might be missed unless we assess balance within the individual subgroups. For example, we might want to allow for the age-exposure relationship in our propensity score model to differ within each subgroup.

It is feasible to implement propensity score weighting separately by subgroup and rescale the weights to produce a new pseudo-population in which all three of these issues are resolved and balance achieved within each subgroup is maintained. We use a similar conceptual approach to the application of stabilized weights and apply weights that depend on the patient's subgroup and exposure status but not on other patient characteristics (see Xu et al.\cite{xu2010} and Hern\'an \& Robins\cite{hernan2020} for more discussion on stabilized weighting). This approach constitutes a stratification in the application of propensity score weighting.

Analysis approaches that incorporate stratification can mitigate confounding due to imbalances in key prognostic factors between treatment/exposure groups and improve the precision of estimates (see Fleming \& DeMets for a discussion in the context of RCTs\cite{cook2007ch1}). This may be particularly important where sample sizes are small or overlap in patient characteristics between exposure groups is limited in propensity score weighting settings. Furthermore, it may also be desirable to ensure that the share of patients in each stratum is unchanged in our weighted pseudo-population so that our marginal estimate of the treatment effect in the overall (all-strata) population will reflect the estimated effect in a patient population with the same stratum proportions as our underlying unweighted data (or we could even re-weight the strata to reflect a new target population to improve transportability\cite{gupta2025}). By fitting separate propensity score models on each stratum, we can easily incorporate adjustment for stratum-specific confounders and modelling considerations (e.g. some clinical prognosis/risk scores or patterns of missing data might be specific to some strata). A final benefit to this approach is that it makes it straightforward to compute valid treatment effect estimates for each of the stratum subgroups in addition to estimates for the overall patient population.

We implement this weighting adjustment through a two-stage approach (which can be modified for estimands other than the ATE):
\begin{enumerate}
\item We multiply the weight $w_{i}$ for each patient $i=1,...,n$ to get a new scaled weight, $w_{i}'$, as follows:
\begin{align*}
w_{i}' = w_{i}\cdot 0.5 \cdot \frac{\sum_{j=1}^{n} w_{j} \cdot 1(s_j = s_i)}{\sum_{j=1}^{n} w_{j} \cdot 1(s_j = s_i, Z_j = Z_i)}\;\;\;\;,
\end{align*}
where $s_i$ is patient $i$'s stratum, and $1(\cdot)$ is an indicator function that equals $1$ if $\cdot$ is true and $0$ otherwise. Or, in words, we multiply each patient's weight by $0.5$ times the ratio of the total weighted patients in that patient's stratum to the total weighted patients in that patient's stratum's exposure group. We add the 0.5 scaling factor so that the total sum of the weights is unchanged after re-weighting but since it is the relative rather than absolute weights that are crucial, this scaling factor could be omitted, a different value could be used, or the outlined approach could even be modified to allow for stabilized weights. 
\item We can then rescale these new weights again so that each stratum's share of the weighted pseudo-population matches the stratum's share in the underlying unweighted population by defining a final set of transformed weights, $w_{i}''$:
\begin{align*}
w_{i}''=w_{i}' \cdot k \cdot \frac{\sum_{j=1}^{n}1(s_j=s_j)}{\sum_{j=1}^{n} w_{j} \cdot 1(s_j = s_i)}\;\;\;\;.
\end{align*} 
\noindent Or, in words, we multiply the $w_{i}'$ by an inflation factor that is the ratio of the number of patients that stratum contributes to the original population relative to the number contributed to the pseudo population and an optional second factor $k=\frac{1}{n}\sum_{j=1}^{n}w_{j}$ that leaves the total weights unchanged after rescaling. Again, $k$ can be dropped since it is a constant and it is only the relative weights that matter. 
\end{enumerate}

As a final step we can fit an outcome regression model on the weighted data using our rescaled $w_{i}''$ propensity score weights. We do this by weighted regression of our outcome $\mathbf{y}$ on the exposure indicator $Z$. We can optionally include additional covariate controls in this model for a quasi-doubly-robust implementation\cite{ho2007}. However, inclusion of additional covariates is not necessary if all confounders are in balance after applying the weights and also changes the estimand from a marginal to a conditional one unless an additional marginalization or `standardization' step is added (see Hern\'an \& Robins\cite{hernan2020} for more discussion of G methods and marginalization approaches and Remiro-Az\'ocar et al.\cite{remiroazocar2025} and Phillippo et al.\cite{phillippo2025} on the implications of the use of marginal vs. conditional causal estimands). 

It is also important when applying propensity score weighting to properly account for the weights when computing standard errors for the treatment effect estimates. Two common approaches are to use robust standard errors (also known as sandwich covariance matrix estimators) or bootstrapping\cite{mansournia2021,austin2016,austin2022}. It is important to note that the precision of effect estimates depends on the variability in the weights rather than their magnitude. Little and Rubin\cite{little2002} provide the following formula for the variance of a sample mean $\bar{Y}$ with constant variance $\sigma^2$ and weights $w$ rescaled so that they sum to $n$:

\begin{align*}
\text{Var}\big[\bar{Y}\big] &= V\Bigg[\frac{1}{n}\sum^n_{i=1}w_iY_i\Bigg] =  \frac{\sigma^2}{n^2}\Bigg[\sum_{i=1}^n w_i^2\Bigg] = \frac{\sigma^2}{n}\Big(1 + CV^2\big[w\big]\Big)\;\;\;\;,
\end{align*}

\noindent where $CV[w]$ is the coefficient of variation of the weights. Note that this is only an approximation as it assumes that the weights are fixed rather than estimated.

Another helpful way to characterize the impact of weighting on the precision of effect estimates is by computing the effective sample size (ESS) of the weighted pseudo-population--the number of unweighted independent observations that would yield equivalent precision\cite{tsd18,golinelli2012}:

\begin{align*}
ESS = \frac{(\sum_{i=1}^n w_i)^2}{\sum_{i=1}^n w_i^2}\;\;\;\;.
\end{align*}

\noindent This formula is also an approximation as it ignores sampling variability in the weights. When there are large imbalances or limited overlap between exposure groups, propensity score weighting can yield extreme weights which reduce both ESS and precision. Consequently, there is a variance-bias trade-off when adjusting for imbalances between exposure groups via propensity score weighting\cite{zubizarreta2015,funk2011,golinelli2012}.

\section{Demonstration}

We demonstrate a basic procedure for applying propensity score weighting that allows for stratification by key clinical subgroups--i.e. where we might be concerned that prognosis, propensity to receive the exposure, or covariate-exposure relationships could differ substantially between these subgroups.

We simulate a toy dataset motivated by an oncology setting with $100$ unexposed patients and $80$ exposed patients and three baseline variables: age, stage (III or IV), and stratum (S1 or S2--these could represent two different tumour types, like melanoma or breast cancer). We simulate the data such that there are imbalances in the age distributions and stage and stratum proportions between exposure groups (see the R code in the appendix for simulation details). 

\begin{table}[h]
\caption{Count of Exposed and Unexposed Patients by Stratum}
\label{crosstab}
\centering
\begin{tabular}[t]{lcc}
\hline
  &   Unexposed & Exposed \\
Stratum & $N=100$ & $N=80$ \\
\hline
S1 & 30 (30.0\%) & 50 (62.5\%) \\
S2 & 70 (70.0\%) & 30 (37.5\%) \\
\hline
\end{tabular}
\end{table}

\begin{figure}[h]
\centering
\caption{Age Distribution by Stratum and Exposure Group for the Example Data}
\label{age_dist}
\includegraphics[scale=1.0]{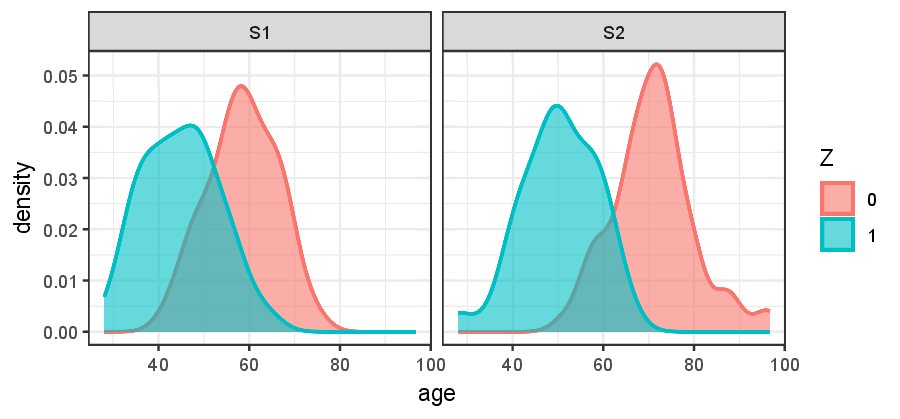}
\end{figure}

For the simulated example data, Table \ref{crosstab} provides a cross-tabulation of the number of patients in each stratum by exposure group (Z). Figure \ref{age_dist} plots the age distributions by stratum and exposure group. 

Table \ref{baltab_un} reports baseline summary statistics prior to adjustment for the example data. We can see that there are notable differences in the mean age between exposed and unexposed (46.9 vs. 67.3 years) and in the proportions of stage IV patients (51.2\% vs. 55.0\%) and stratum S2 patients (37.5\% vs. 70.0\%). The table also reports standardized mean differences (SMD) between the exposed and unexposed groups for these baseline characteristics.

\begin{table}[h]
\caption{Pre- and Post-Adjustment Balance (Un-Stratified)}
\label{baltab_un}
\centering
\begin{tabular}[t]{lrrrcrrr}
\hline
  & \multicolumn{3}{c}{Unadjusted} & & \multicolumn{3}{c}{Adjusted}\\
  \cline{2-4}\cline{6-8}
  & Unexposed & Exposed & SMD & & Unexposed & Exposed & SMD\\  
  & $N=100$ & $N=80$ &  & & $\text{ESS}=38.2$ & $\text{ESS}=43.8$ & \\
\hline
Age (Mean) & 67.257 & 46.943 & -2.139 & & 60.647 & 51.468 & -0.966\\
Stage IV (Prop.) & 0.550 & 0.512 & -0.075 & & 0.493 & 0.533 & 0.080\\
stratum S2 (Prop.) & 0.700 & 0.375 & -0.689 & & 0.540 & 0.429 & -0.236\\
\hline
\end{tabular}
\end{table}

For the purpose of comparison, we first use a typical propensity score weighting approach to adjust for the imbalances between the exposure groups. We estimate the propensity scores using a logistic regression model which regresses the exposure indicator, $Z$, on age, a binary indicator for stage IV, a binary indicator for stratum S1, and age-stratum and stage-stratum interaction terms (to account for the fact that both the age distribution and stage IV proportions differ within each stratum. We then compute our ATE weights and an analogous set of summary statistics in the weighted population to assess post-weighting balance.

Table \ref{baltab_un} reports the summary statistics after re-weighting the data based on the ATE weights (the `adjusted' summaries). We can see that the SMD for age has fallen in magnitude from -2.139 to -0.966 and, for the stratum S2 proportions, from -0.689 to -0.236. The change in the stage IV proportion is more modest, with the SMD going from -0.075 to 0.080. Note that due to the small sample sizes and large imbalances in baseline characteristics between exposure groups in the unadjusted data, we are unable to achieve good balance even after applying propensity score weighting. Furthermore, we can see that, not only is there a notable imbalance remaining between exposure groups in the proportion of patients in each stratum (SMD = -0.236), but the share of stratum S2 patients in the adjusted data (49.1\%) is lower than the share in the underlying data (55.6\%)--meaning that weighted estimates of the effect of exposure will reflect the effect in a target population in which stratum S2 patients are under-weighted.

We now consider the application of the proposed stratified propensity score weighting approach. For this method we fit separate propensity score models on each stratum which we then use to compute ATE weights for each stratum as usual. We then apply the re-weighting approach outlined above to compute our adjusted weights $w''$. Lastly we apply these $w''$ weights to construct our weighted pseudo-population and assess post-adjustment balance between exposure groups.

Table \ref{baltab_s} reports both unadjusted and adjusted balance statistics using the stratified approach. In contrast to the earlier unstratified approach, we see that there is now perfect balance in the Stratum S2 proportions between exposure groups (SMD = 0.000). Additionally, the overall share of Stratum S2 patients under this weighting is now 55.6\%--which matches the share in the original unweighted data. Nevertheless, it should be noted that residual imbalances remain for the age and stage variables.

\begin{table}[h]
\caption{Pre- and Post-Adjustment Balance (Stratified)}
\label{baltab_s}
\centering
\begin{tabular}[t]{lrrrcrrr}
\hline
  & \multicolumn{3}{c}{Unadjusted} & & \multicolumn{3}{c}{Adjusted}\\
  \cline{2-4}\cline{6-8}
  & Unexposed & Exposed & SMD & & Unexposed & Exposed & SMD\\  
  & $N=100$ & $N=80$ &  & & $\text{ESS}=39.6$ & $\text{ESS}=34.6$ & \\
\hline
Age (Mean) & 67.257 & 46.943 & -2.139 & & 60.888 & 52.307 & -0.903\\
Stage IV (Prop.) & 0.550 & 0.512 & -0.075 & & 0.491 & 0.508 & 0.034\\
Stratum S2 (Prop.) & 0.700 & 0.375 & -0.689 & & 0.556 & 0.556 & 0.000\\
\hline
\end{tabular}
\end{table}

It is also important to consider balance within each stratum--especially when explicitly performing a stratified analysis by important clinical factors. Table \ref{baltab_s_within} reports a similar set of balance summaries separately by stratum. We can see that, not only is age imbalanced (especially within stratum S2) but also that a notable imbalance in the stage IV proportion is now visible within stratum S1 (61.5\% vs. 55.6\%).

\begin{table}[h]
\caption{Within-Stratum Pre- and Post-Adjustment Balance (Stratified)}
\label{baltab_s_within}
\centering
\begin{tabular}[t]{lrrrcrrr}
\hline
  & \multicolumn{3}{c}{Unadjusted} & & \multicolumn{3}{c}{Adjusted}\\
  \cline{2-4}\cline{6-8}
  & Unexposed & Exposed & SMD & & Unexposed & Exposed & SMD\\  
  & $N=100$ & $N=80$ &  & & $\text{ESS}=39.6$ & $\text{ESS}=34.6$ & \\
\hline
\textbf{\textit{Stratum S1 ($N=80$)}} & & & & & & & \\ 
\;\;\;\;\;Age (Mean) & 58.480 & 44.952 & -1.711 & & 52.082 & 48.623 & -0.438\\
\;\;\;\;\;stage IV (Prop.) & 0.767 & 0.620 & -0.322 & & 0.555 & 0.615 & 0.133 \vspace*{0.07in} \\
\textbf{\textit{Stratum S2 ($N=100$)}} & & & & & & & \\
\;\;\;\;\;Age (Mean) & 71.019 & 50.261 & -2.407 & & 67.934 & 55.254 & -1.470\\
\;\;\;\;\;stage IV (Prop.) & 0.457 & 0.333 & -0.255 & & 0.440 & 0.422 & -0.037\\
\hline
\end{tabular}
\end{table}

\section{Discussion}

In certain clinical research settings we are interested in using observational data to estimate marginal treatment effects in patient populations which include multiple clinical strata in which prognosis and prognostic factors can differ substantially. When using a method like propensity score weighting to adjust for potential confounders in the estimation of the marginal treatment effect, we may want to prioritize achieving balance for the strata proportions between exposure groups. We have outlined a potential approach that practitioners can use to achieve this goal. The approach applies propensity score weighting within each stratum and, after applying a scaling factor to the weights, marginalizes over the strata to get a treatment effect estimate for the overall population.

Another key feature of this approach is the ability to fit different propensity score models for the individual strata. This can be advantageous when there are clinical variables that are specific to some strata (e.g. cancer stage for cancer treatments, or immune status for prophylactic antibiotics), or where the the prognostic importance of some variables or their impact on treatment uptake is expected to differ between strata. Application of this method may be particularly relevant where sample sizes are limited and achieving balance in key clinical subgroups needs to be prioritized. 

We recommended that practitioners using this approach conduct propensity score weighting diagnostics within each stratum, in addition to the overall combined weighted pseudo-population. This should include (a) assessing the distribution of baseline characteristics between exposure groups (e.g. comparing summary statistics like standardized mean differences between weighted exposure groups for each potential confounder\cite{austin2009}), (b) assessing the distribution of propensity scores to diagnosis potential non-overlap (positivity) issues, (c) inspecting the distribution of weights for extreme values, (d) computing effective sample sizes (ESS) for each exposure group, and (e) other context-specific diagnostics as appropriate (e.g. weighted log-cumulative hazard plots to assess the proportional hazards assumption for Cox proportional hazards survival models\cite{latimer2011}). 

Care should also be given to the specification of both the exposure and outcome regression models--both in terms of the identification of potential confounders and in terms of their parameterization. See VanderWeele\cite{vanderweele2019} for an overview of good practices in confounder selection and Harrell\cite{harrell2015ch2} for a discussion of parametric modelling approaches. Where sample sizes are limited, exposures/non-exposures are rare, or outcome events are rare for binary or survival outcomes, special care should be taken in determining appropriate model complexity (see Harrell\cite{harrell2015ch4} for further discussion). Ho et al.\cite{ho2007} also note that a quasi-doubly-robust approach can be implemented by including potential confounders as additional covariates in the outcome regression model (see Funk et al.\cite{funk2011} for an overview of doubly-robust estimation). Note, however, that the inclusion of additional covariates in the outcome regression model will impact the estimand and interpretation of the treatment effect estimated by the model as it will no longer be a marginal but a conditional average treatment effect.

Implementing these methods within a broader target trial emulation framework is advisable to further mitigate other sources of bias beyond confounding\cite{hernan2016,hernan2022}. Target trial emulation is seeing increased uptake\cite{scola2023,zuo2023} as it provides a structured framework that ensures careful definition of the exposure, timing at which baseline characteristics (potential confounders) are measured, a well-defined time-zero or index date from which outcomes are measured, and a well-defined estimand (see Remiro-Az\'ocar et al.\cite{remiroazocar2025} and ICH E9 (R1)\cite{ichE9R1} for a thorough discussion of estimands). 

Lastly, care needs to be taken to ensure that an appropriate method is used to compute standard errors that account for the weighting applied in the outcome regression model (or weighted non-parametric estimates). Appropriate approaches may include the use of robust standard errors (also know as sandwich covariance matrix estimators), or bootstrapping (where the propensity score estimation, weighting, and outcome regression steps are repeated for each bootstrapped sample)\cite{austin2016,austin2022}.

Similar methods for computing marginal estimates of treatment effects with stratification can also be implemented for matching, covariate balancing weighting, or standardization/parametric g-computation. For covariate balancing weights the procedure will be similar apart from the replacement of the propensity score model with an optimization problem that balances covariates. For parametric g-computation a separate outcome regression can be fit for each stratum with a standardization procedure. Implementation of a similar procedure for matching will be feasible in some cases but may be more difficult to implement in practice as the use of calipers, unmatched cases, and specific features of the choice of matching method may complicate the implementation (see Stuart\cite{stuart2010} for a review of matching methods). 

The outlined approach is subject to the same limitations as standard propensity score approaches. Several assumptions must hold for the validity of the method (or, in practice, at least approximately hold). The exposure status of one patient must not impact the outcome or exposure status of other patients (the SUTVA assumption). The method requires that all confounders be observed (the ignorability assumption) and that the propensity score model incorporating these confounders is correctly specified. Lastly, where there is limited overlap in the distribution of confounders between the exposed and unexposed groups (positivity violations or near-violations), it may be difficult to achieve good balance in baseline characteristics between exposure groups after applying weighting, effective sample sizes may be low, and results may be sensitive to a small number of observations with extreme weights.  

\section{Conclusion}

This article aimed to outline a propensity score weighting approach which incorporates stratification by clinical subgroups. The method allows for practitioners to ensure balance between exposure groups in the shares of patients in each stratum after applying the propensity score weights. It also allows for separate propensity score models to be estimated for each stratum. Outcomes can then be compared between exposure groups for the weighted pseudo-population constructed using adjusted propensity score weights. Practical guidance on the application of propensity score methods with stratification was also provided. Propensity score weighting provides a useful approach for mitigating confounding when estimating causal effects using observational data. Allowing for stratification in the propensity score weighting analysis further extends the utility of this method when estimating average treatment effects where there is substantial heterogeneity across subpopulations of patients.

\printbibliography

\section*{Appendix: R Code for Demonstration}

\begin{verbatim}
library(tidyverse)
library(cobalt)
library(knitr)

#set random seed
set.seed(21082025)

#parameters for simulating baseline characteristic data
n <- c(30, 50, 70, 30)
mu_age <- c(60, 45, 70, 50)
sigma_age <- c(8, 8, 8, 8)
prop_stageIV <- c(0.6, 0.55, 0.45, 0.43)

#simulate baseline characteristics data
dat <- data.frame(group = rep(1:4, n),
                  stratum = factor(rep(c("S1", "S2"), c(n[1] + n[2], n[3] + n[4]))),
                  Z = rep(c(0, 1, 0, 1), n))
dat$age <- rnorm(sum(n), mean = mu_age[dat$group], sd = sigma_age[dat$group])
dat$stage <- factor(1 + rbinom(sum(n), 1, prop_stageIV[dat$group]),
                    labels = c("III", "IV"))
dat$stage_IV <- 1 * (dat$stage == "IV")
dat$stratum_S2 <- 1 * (dat$stratum == "S2")

#show distribution of age by exposure and stratum
cairo_ps("fig1.eps", width = 6, height = 2.8)
ggplot(dat %>% mutate(Z = factor(Z)), aes(x = age, fill = Z, color = Z)) + 
  geom_density(linewidth = 0.9, alpha = 0.6) + facet_wrap(~ stratum) + theme_bw() 
dev.off()

#cross-tabulation of stratum and exposure
tmp <- table(dat$stratum, dat$Z)
print(tmp)
prop.table(tmp, margin = 2)

#function to compute ATE weights for IPTW via logistic regression
get_ATE_weights <- function(formula, data) {
  fit <- glm(formula, data = data, family = binomial(link = "logit"))
  ps <- predict(fit, type = "response")
  w <- (data$Z / ps) + ((1 - data$Z) / (1 - ps))
  return(w)
}

#compute un-stratified (u) ATE weights
dat$w_u <- get_ATE_weights(Z ~ age + stage + stratum + age:stratum + stage:stratum, dat)

#compare balance before and after un-stratified adjustment
tmp <- bal.tab(model.matrix(~ -1 + age + stage_IV + stratum_S2, data = dat), 
               treat = dat$Z, weights = dat$w_u,
               continuous = "std", binary = "std",
               un = T, disp = c("means"), s.d.denom = "pooled")
kable(tmp$Balance, digits = 3, , format = "latex",
      caption = "Pre- and Post-Adjustment Balance (Un-Stratified)")
tmp$Observations

#get separate ATE weights by stratum
S1 <- dat %>% filter(stratum == "S1")
S2 <- dat %>% filter(stratum == "S2")
S1$w <- get_ATE_weights(Z ~ age + stage, S1)
S2$w <- get_ATE_weights(Z ~ age + stage, S2)
dat <- rbind(S1, S2)

#create lookup tables to compute stratified weights
lk_w_s <- dat %>% group_by(stratum) %>% summarise(val = sum(w))
lk_w_sz <- dat %>% group_by(stratum, Z) %>% summarise(val = sum(w))
lk_n_s <- dat %>% group_by(stratum) %>% summarise(val = n())

#compute the w' and w'' weights
wsum <- sum(dat$w)
dat$wp <- numeric(nrow(dat))
dat$wpp <- numeric(nrow(dat))
for (i in 1:nrow(dat)) {
  dat$wp[i] <- dat$w[i] * 0.5 * lk_w_s[lk_w_s$stratum == dat$stratum[i],]$val /
    lk_w_sz[lk_w_sz$stratum == dat$stratum[i] & lk_w_sz$Z == dat$Z[i],]$val
  dat$wpp[i] <- dat$wp[i] * (wsum / lk_w_s[lk_w_s$stratum == dat$stratum[i],]$val) *
    (lk_n_s[lk_n_s$stratum == dat$stratum[i],]$val / nrow(dat))
}

#compare balance before and after stratified adjustment
tmp <- bal.tab(model.matrix(~ -1 + age + stage_IV + stratum_S2, data = dat), 
               treat = dat$Z, weights = dat$wpp,
               continuous = "std", binary = "std",
               un = T, disp = c("means"), s.d.denom = "pooled")
kable(tmp$Balance, digits = 3, format = "latex",
      caption = "Pre- and Post-Adjustment Balance (Stratified)")
tmp$Observations

#compute a similar set of within-stratum summaries
t1 <- bal.tab(model.matrix(~ -1 + age + stage_IV, 
                            data = dat %>% filter(stratum == "S1")), 
              treat = dat$Z[dat$stratum == "S1"], 
              weights = dat$wpp[dat$stratum == "S1"],
              continuous = "std", binary = "std",
              un = T, disp = c("means"), s.d.denom = "pooled")
t2 <- bal.tab(model.matrix(~ -1 + age + stage_IV, 
                           data = dat %>% filter(stratum == "S2")), 
              treat = dat$Z[dat$stratum == "S2"], 
              weights = dat$wpp[dat$stratum == "S2"],
              continuous = "std", binary = "std",
              un = T, disp = c("means"), s.d.denom = "pooled")
kable(rbind(t1$Balance, t2$Balance), digits = 3, format = "latex",
      caption = "Within-Stratum Pre- and Post-Adjustment Balance (Stratified)")

#compute stratum shares before and after weighting for the un-stratified and 
#   stratified approaches
mean(dat$stratum_S2)
sum(dat$w_u * dat$stratum_S2) / sum(dat$w_u)
sum(dat$wpp * dat$stratum_S2) / sum(dat$wpp)
\end{verbatim}

\end{document}